\documentclass[aps,prb,twocolumn,superscriptaddress,english]{revtex4-2}
\usepackage[utf8]{inputenc}

\usepackage{amsmath}
\usepackage{amssymb}
\usepackage{graphicx}
\usepackage{xcolor}
\usepackage{siunitx}
\usepackage{lipsum}
\usepackage{orcidlink}
\usepackage{gensymb}
\usepackage{lineno}
\usepackage{float}

\usepackage{hyperref}
\usepackage{orcidlink}
\usepackage{xr-hyper}

\newcommand*{\addFileDependency}[1]{
\typeout{(#1)}
\@addtofilelist{#1}
\IfFileExists{#1}{}{\typeout{No file #1.}}
}\makeatother

\makeatletter
\def\@fnsymbol#1{\ensuremath{\ifcase#1\or \dagger\or *\or \ddagger\or
   \mathsection\or \mathparagraph\or \|\or **\or \dagger\dagger
   \or \ddagger\ddagger \else\@ctrerr\fi}}
\makeatother

\begin{document}

\author{Loan Renaud}\affiliation{Laboratoire de physique de L'École normale supérieure de Paris, CNRS, ENS \& Université PSL, Sorbonne Université, Université de Paris, 75005 Paris, France}
\author{Tomasz Poreba}\affiliation{Laboratory for Quantum Magnetism, Institute of Physics, \'{E}cole Polytechnique F\'{e}d\'{e}erale de Lausanne, CH-1015 Lausanne, Switzerland}
\author{Richard Gaal}\affiliation{Laboratory for Quantum Magnetism, Institute of Physics, \'{E}cole Polytechnique F\'{e}d\'{e}erale de Lausanne, CH-1015 Lausanne, Switzerland}
\author{A. Marco Saitta}\affiliation{Laboratoire de physique de L'École normale supérieure de Paris, CNRS, ENS \& Université PSL, Sorbonne Université, Université de Paris, 75005 Paris, France}
\affiliation{Institut Universitaire de France (IUF)}
\author{Michele Casula}\affiliation{Institut de Min\'{e}ralogie, de Physique des Mat\'{e}riaux et de Cosmochimie (IMPMC), Sorbonne Universit\'{e}, CNRS UMR 7590, MNHN, 4, place Jussieu, Paris, France}
\author{Livia Eleonora Bove}\affiliation{Laboratory for Quantum Magnetism, Institute of Physics, \'{E}cole Polytechnique F\'{e}d\'{e}erale de Lausanne, CH-1015 Lausanne, Switzerland}\affiliation{Institut de Min\'{e}ralogie, de Physique des Mat\'{e}riaux et de Cosmochimie (IMPMC), Sorbonne Universit\'{e}, CNRS UMR 7590, MNHN, 4, place Jussieu, Paris, France}
\affiliation{Dipartimento di Fisica, Sapienza Universit\`a di Roma, Piazzale Aldo Moro 5, 00185 Roma, Italy}

\title{Negative Thermal Expansion in Cubic Ice: A Collective Quantum Effect of the hydrogen-bond network}

\begin{abstract}

We report neutron powder diffraction measurements and path-integral molecular dynamics simulations of stacking-disorder-free cubic ice I$_c$, produced by topotactic degassing of C2 hydrogen hydrate. Across the cryogenic stability range, ice I$_c$ exhibits a density maximum near 70 K, closely matching that of hexagonal ice I$_h$ despite their different long-range stacking sequences. Negative thermal expansion in ice I is therefore not specific to hexagonal stacking, but arises from the shared open tetrahedral hydrogen-bond network. Simulations with the MB-pol potential quantitatively reproduce the experimental anomaly only when nuclear quantum effects are included. The density maximum coincides, within the temperature resolution, with maximal anisotropy of the proton quantum distribution. Neutron-derived displacement parameters independently reveal a strongly enhanced transverse proton displacement, while phonon calculations identify low-frequency transverse modes with the most negative Grüneisen parameters. Together, these results establish the negative thermal expansion of ice I as a collective quantum effect governed by nuclear statistics and the dynamics of the hydrogen-bond network.


\end{abstract}

\maketitle

Negative thermal expansion (NTE) --- the contraction of a solid upon heating --- occurs in open
framework materials when low-lying vibrational modes with negative Gr\"{u}neisen parameters
are thermally populated \cite{Evans1999, Barrera2005, Strssle2004}. In many known NTE systems, including oxide
perovskites such as ZrW$_2$O$_8$, cyanide frameworks and zeolites \cite{Goodwin2008, Lightfoot2001, Fang2014, Evans1996}, 
this behaviour can often be captured within classical lattice-dynamical or molecular-dynamics (MD) descriptions  \cite{Schick2016}.
Hexagonal ice I$_h$, however, is exceptional. Its lattice contracts below $\sim$70~K
\cite{Rttger1994, Tanaka2001, Herrero2011, Pamuk2012, Buckingham2018}, yet this anomaly disappears when nuclear motion is treated classically. Classical MD predicts no density maximum, whereas path-integral molecular dynamics (PIMD), which incorporates nuclear quantum effects (NQE), quantitatively reproduces the experimental behavior \cite{Herrero2011,Eltareb2023}. The NTE of ice I$_h$ is therefore a nuclear quantum phenomenon \cite{Ceriotti2016}, but whether it originates from local quantum fluctuations arising from individual O--H bonds, or from the collective dynamics of the hydrogen-bond network remains unclear.
 
Distinguishing these two pictures requires a system that preserves the local tetrahedral environment while changing the long-range stacking topology. Cubic ice (I$_c$), the metastable
polymorph with diamond-cubic oxygen sublattice ($Fd\bar{3}m$  space group), provides exactly such a testbed:
I$_c$ and I$_h$ are isocompositional and share esentially the same local tetrahedral coordination and closely similar O--H
bond length and
O--O
distance, while differing in
long-range stacking sequence and phonon dispersion. Until recently, however, this comparison was impossible because nominal cubic ice typically contained stacking disorder (I$_{sd}$) \cite{Malkin2014}, whose hexagonal faults perturb the lattice dynamics \cite{Carr2014}.
Stacking-disorder-free ice I$_c$ became available only in 2020 \cite{Komatsu2020, delRosso2020, delrosso2026}. The C2-hydrate route yields a cubic phase that remains sufficiently stable throughout the temperature interval containing the NTE anomaly.

Here we combine neutron powder diffraction on stacking-disorder-free ice I$_c$ with classical MD and PIMD using the MB-pol interatomic potential \cite{Babin2014, Babin2013, Medders2014, Riera2023, Paesani2016,Reddy2016} to determine how NTE depends on crystal topology and to uncover its microscopic origin. We find that ice I$_c$ exhibits the same density maximum near 70 K as ice I$_h$, demonstrating that NTE is an intrinsic property of the open tetrahedral hydrogen-bond network and is largely insensitive to the stacking sequence. To unveil the role of the network, we analyse the spread of the simulated quantum paths as a function of temperature. We find that the maximum anisotropy of the proton distribution coincides with the temperature of the maximum contraction of the crystal, directly linking the proton quantum state to the macroscopic lattice response.
This anisotropy is supported by neutron-diffraction data
suggesting that the transverse phonon modes play a key role in the NTE. Consistently, these transverse modes exhibit the most negative Grüneisen parameters.
This behavior is a signature of a collective quantum effect, one that can only be captured if the crystal environment itself is treated as an ensemble of quantum particles.

Stacking-disorder-free cubic ice I$_c$ was prepared by topotactic degassing of C$_2$ hydrogen hydrate \cite{Antonov2009} following the procedure of Komatsu \textit{et al.} \cite{Komatsu2020}. C$_2$ hydrogen hydrate was synthesized in a Paris--Edinburgh press and decompressed under liquid-nitrogen conditions to preserve the cubic oxygen framework while allowing complete desorption of the guest D$_2$ molecules. The recovered sample was transferred under cryogenic conditions to the D20 neutron diffractometer, where diffraction patterns were collected between 50 and 200 K (see SM \cite{SM}).

\begin{figure}
    \centering
    \includegraphics[width=\linewidth]{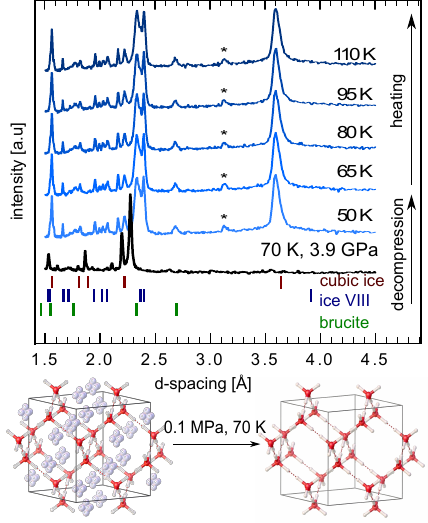}
    \caption{Representative neutron diffractograms of cubic ice (blue lines) produced by 
    decompression of C2 hydrogen hydrate (black line) at low temperature. Weak Bragg reflections 
    from unreacted MgD$_{2}$ are marked with asterisks. Bottom inset: release of guest D$_2$ from the C2 structure upon decompression at 70 K, resulting in the formation of cubic ice.}
    \label{fig:fig1}
\end{figure}

Figure \ref{fig:fig1}(a) compares the neutron diffraction patterns of C$_2$ hydrogen hydrate before degassing and the recovered sample at 50 K. The disappearance of the diffuse scattering associated with disordered guest D$_2$ and the emergence of the intense I$_c$ (111) reflection provide clear signatures of the topotactic transformation. Rietveld refinement yields a D$_2$ occupancy of 0.00(4) at the 48$f$ guest site, confirming complete hydrogen removal. The recovered major phase is identified as stacking-disorder-free cubic ice I$_c$ (\textit{Fd$\bar{3}$m}), with no detectable I$_h$ contribution within the sensitivity of the neutron data. Minor amounts of residual ice VIII, brucite, and traces of unreacted MgD$_2$ arise from the synthesis but do not interfere with the refinement of the cubic phase.

The cubic phase remains stable over the 50-200 K interval used for the thermal-expansion analysis. Above approximately 200 K, stacking disorder progressively develops, followed by transformation towards ice I$_h$, preventing an unambiguous determination of the cubic lattice parameter. We therefore extracted the I$_c$ lattice parameter by Le Bail refinement in Topas V6 software \cite{LeBail_2005, Coelho2018} at 15 K intervals between 50 and 200 K (see SM \cite{SM}).

\begin{figure}
    \centering
    \includegraphics[width=1\linewidth]{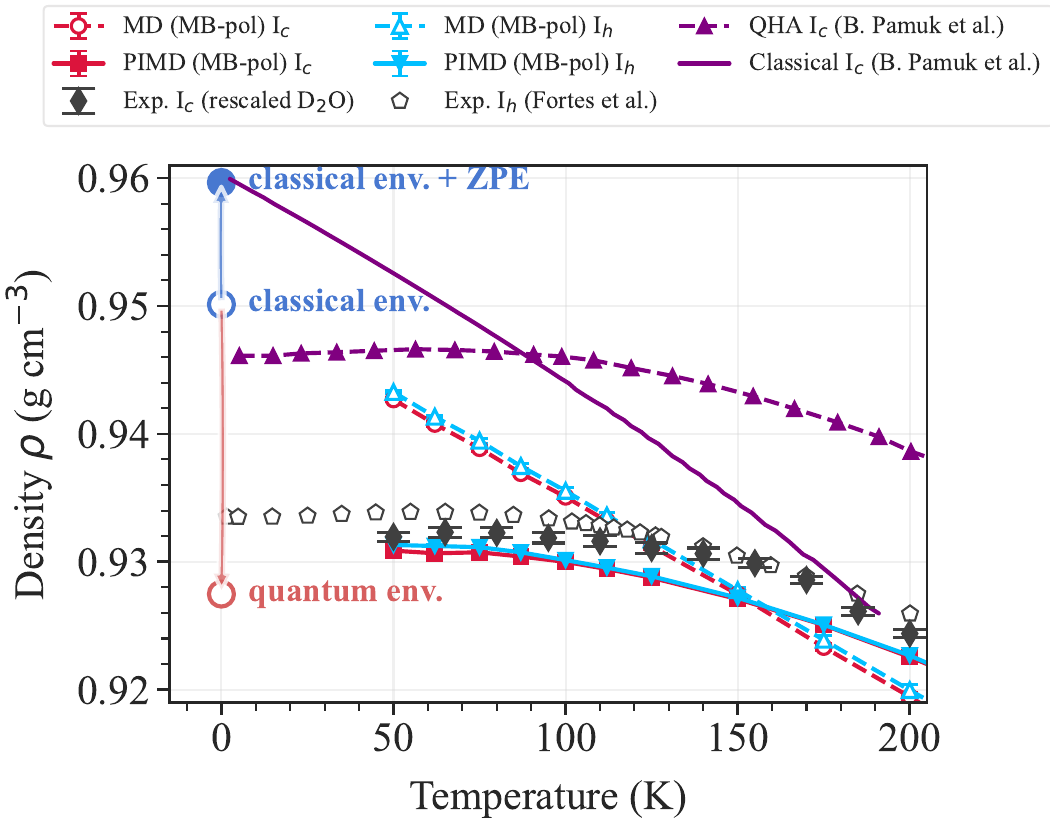}
    \caption{Density of cubic (red) and hexagonal (blue) ice as a function of temperature,
from classical MD (dashed lines), PIMD (solid lines), and experimental measurements (data points) \cite{Fortes2018}.
    At $T=0$~K, open blue (red) markers
    indicate densities derived from 
    minimizing the lattice potential generated by a classical (quantum) environment;
    see Fig.~\ref{fig:fig4} and SM \cite{SM}.
    The solid blue marker includes the zero-point energy (ZPE) correction,
obtained by solving 
the
    Schr\"{o}dinger equation for a proton in the lattice potential. We also report  QHA and classical MD results from \cite{Pamuk2012} (purple points).
    }
    \label{fig:fig2}
\end{figure}

Experimental and simulated densities of I$_c$ and I$_h$ are plotted in Figure~\ref{fig:fig2} as a function of temperature at ambient pressure. The experimental I$_c$
density was derived from D$_2$O measurements using a mass-based scaling factor, a procedure
validated in the Supplemental Material (SM \cite{SM}). The two polymorphs display nearly identical density
profiles across most of the temperature range. A slight difference is measured at low temperatures, with I$_h$ maintaining a marginally
higher density, reflecting the subtle influence of the different stacking sequences on the lattice
dynamics. 
The close proximity of the density maxima in I$_c$ and I$_h$ is itself informative. The small difference indicates that stacking leaves a measurable imprint on the lattice dynamics, but acts only as a secondary modulation of a mechanism common to both ice-I polytypes.
The similarity in both magnitude and characteristic temperature shows that the mechanism is associated with the tetrahedral hydrogen-bond network shared by the ice-I polytypes, rather than with a specific stacking sequence.

The simulations reveal the quantum origin of the NTE. Classical MD predicts a monotonically
increasing density upon cooling, with no density maximum for either polymorph~\cite{Pamuk2018,
Eltareb2023}. PIMD, which accounts for NQE through its ring-polymer representation, 
quantitatively reproduces the density maximum and agrees with experiment across the
entire temperature range, with a maximum relative error of $\sim$0.2\% — an order of magnitude
smaller than previous estimates on I$_h$ using the q-TIP4P/f potential~\cite{Habershon2009, Eltareb2023}. Because the classical and path-integral simulations use the same MB-pol potential, their qualitative difference cannot be attributed simply to the underlying potential-energy surface; it arises from the treatment of nuclear statistics. Above 150~K, PIMD also yields a systematically
higher density than classical MD, consistent with earlier QHA~\cite{Pamuk2018} and PIMD
studies~\cite{Cheng2019} showing that nuclear quantum effects influence the equilibrium volume beyond the NTE regime.

\begin{figure}
    \centering
    \includegraphics[width=1\linewidth]{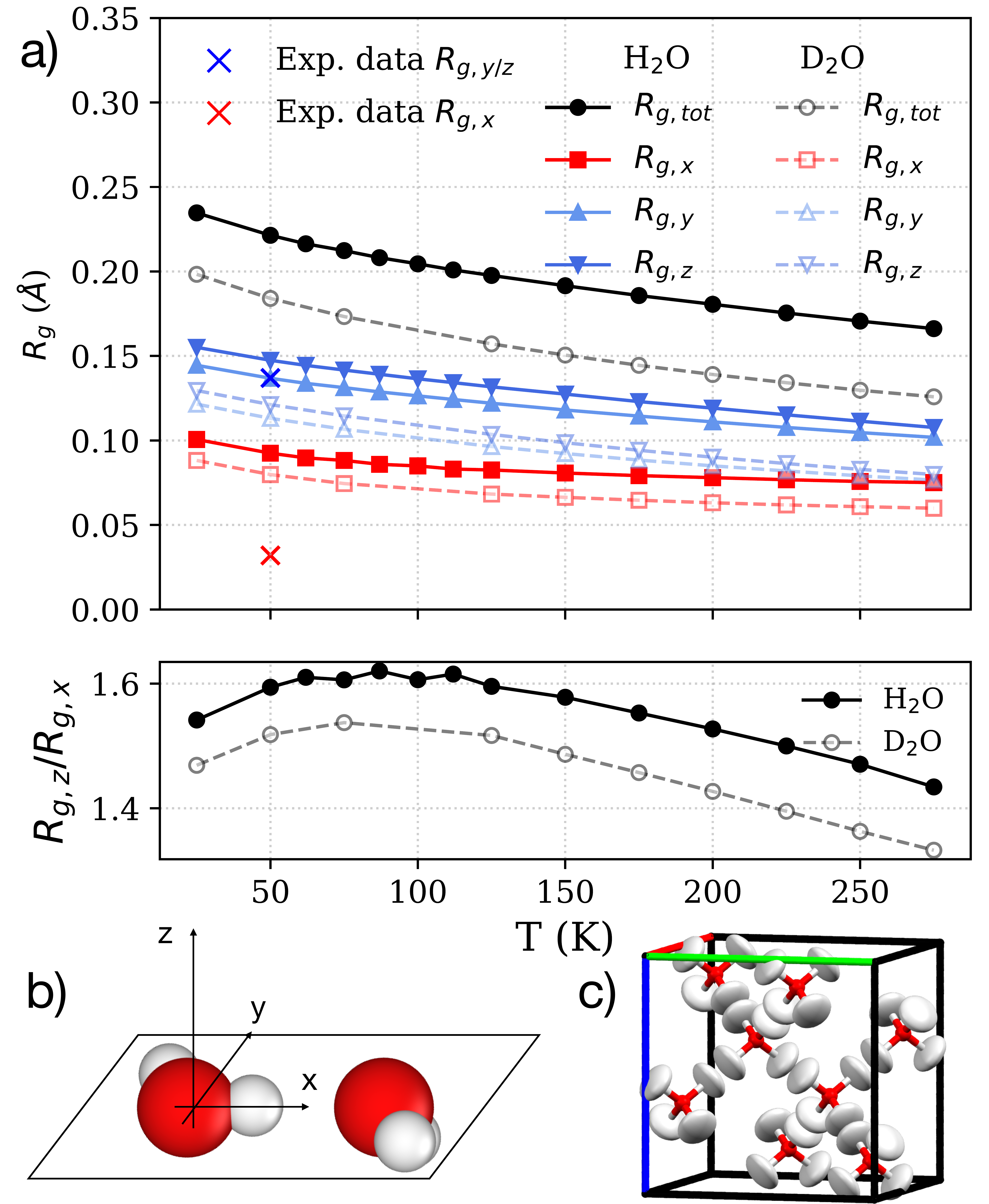}
    \caption{(a, top): Temperature-dependent gyration radius $R_g$ and its components $R_{g,\alpha}$ for $\alpha \in \{x,y,z\}$ (Eq.~\ref{eq:gyration})
    computed for H$_2$O (solid symbols) and
    D$_2$O (open symbols) in ice I$_c$. 
Red (blue) cross at 50 K is the longitudinal ($x$) and transverse ($y/z$) value from the experimental anisotropic thermal displacement.
    (a, bottom):
    Transverse-over-longitudinal anisotropy ratio ($R_{g,z}/R_{g,x}$) from PIMD. (b): 
    Representation
    of the local
    molecular frame, 
    following Ref.~\cite{Eltareb2023}. (c): Experimental
    anisotropic thermal displacement ellipsoids 
    in the I$_c$ conventional cubic cell
    at 50~K 
    from powder neutron diffraction. The ellipsoids represent
    50\% probability level isosurfaces.}
    \label{fig:fig3}
\end{figure}

To study the manifestation of NQE in I$_c$, we dissect the quantum proton contribution from trivial thermal fluctuations by computing
the average gyration radius $R_g$ of the PIMD ring polymers. This quantity, resolved in local Cartesian components $\alpha \in \{x, y, z\}$, is defined as \cite{Tuckerman2023}
\begin{equation}
    R_{g,\alpha}^2 = \left \langle \frac{1}{n_b} \sum_{i=1}^{n_b}(r_{c,\alpha}-r_{i,\alpha})^2 \right \rangle,
    \label{eq:gyration}
\end{equation} 
where $r_{i,\alpha}$ and $r_{c,\alpha}$ are respectively the bead and centroid coordinates of a proton,   $\langle \rangle$ indicates the time-averaged over all the protons in the simulation supercell, $n_b$ is the number of beads of the ring polymers. The Cartesian components are computed with respect to a local instantaneous frame which depends on the centroid positions of each H$_2$O molecule in the crystal \cite{Eltareb2023}, where the $x$ axis lies along the O--H bond, the $z$ axis is perpendicular to the molecular plane, and the $y$ axis forms the remaining orthogonal direction, as shown in Fig.~\ref{fig:fig3}(b).
The results in Fig.~\ref{fig:fig3}(a) show a pronounced quantum anisotropy: the proton is approximately 1.5 times more delocalized in the transverse directions ($R_{g,y}$, $R_{g,z}$) than along the covalent O--H bond ($R_{g,x}$), consistent with earlier findings for I$_h$~\cite{Eltareb2023}. However, our analysis reveals a critical new feature: while the absolute magnitude of $R_g^2 \equiv \sum_\alpha R_{g,\alpha}^2$ and all its $\alpha$ components decrease with temperature,
the anisotropy ratio $R_{g,z}/R_{g,x}$ exhibits a broad maximum that coincides, within the temperature resolution, with the density maximum reported in~Fig. \ref{fig:fig2}. Our observation 
supports 
a direct link between the anisotropic proton quantum distribution and the macroscopic NTE. The isotope dependence reinforces this picture: H$_2$O shows a systematically larger $R_g$ than D$_2$O at all temperatures, consistent with its lighter mass, but leaves the anisotropy ratio almost unchanged, in qualitative agreement with the weak isotope dependence of NTE found in our calculations (see SM \cite{SM}).
The experimental anisotropic thermal displacement ellipsoids determined from neutron powder
diffraction via Rietveld refinement \cite{Rietveld_1969} at 50~K provide independent evidence for a strongly anisotropic proton distribution, with substantially larger displacements transverse to the O-H bond (Fig.~\ref{fig:fig3}(c)). The transverse displacement amplitude $\sim$0.137~\AA{} is close to the PIMD value and smaller than the 0.184~\AA\ reported in earlier refinement by Kuhs \cite{Kuhs1983}. The longitudinal
amplitude, $\sim$0.032~\AA, is substantially smaller than the corresponding ring-polymer component, yielding an even stronger experimental anisotropy. Because diffraction displacement parameters cannot be identified directly with ring-polymer gyration radii, the comparison should be considered qualitative.

\begin{figure*}[th!]
    \centering
    \includegraphics[width=0.9\textwidth]{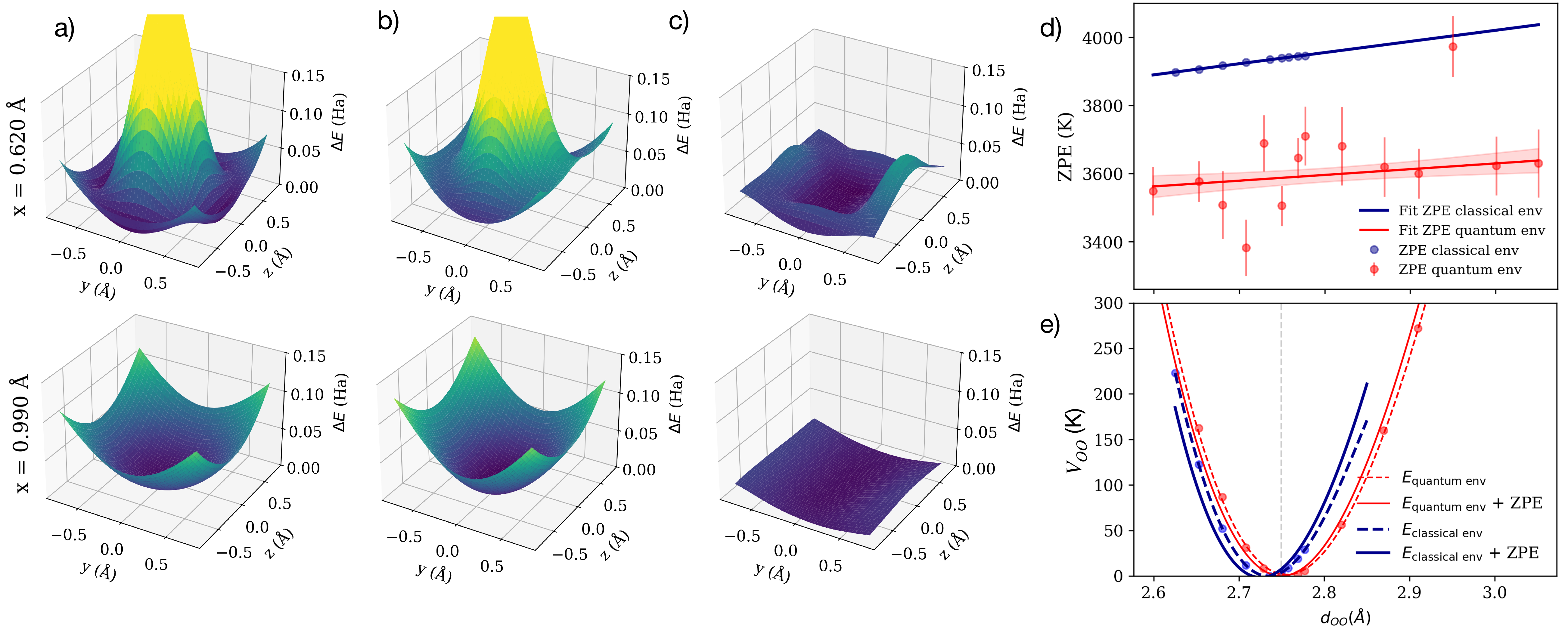}
    \caption{Proton in the I$_c$ environment. Slices of the interpolated PES in the molecular local frame along the O--H bond ($x$ axis) for classical (a) and quantum (b) environments. The PES is shown at the proton position in the repulsive region ($x=0.620$\AA) and around equilibrium ($x=0.990$\AA). (c): PES difference between (a) and (b). (d): Proton ZPE as a function of the O--O distance, $d_\mathrm{OO}$, for both models. (e): Interatomic potential $V_{OO}=V_{OO}(d_{OO})$.
    Dashed and solid lines represent the bare $V_{OO}$ and the one including proton ZPE, respectively.}
    \label{fig:fig4}
\end{figure*}

To quantify the link between the proton quantum state and the macroscopic lattice response, we studied how the effective O--O interatomic potential $V_{OO}$, dictating the cell volume expansion, is renormalized by this non-trivial proton dynamics \cite{Mouhat2023, Min2025}. 
We therefore constructed the problem of a
single-proton embedded in the I$_c$ crystal environment, and solved it at different O--O distances $d_{OO}$. 
By this construction, we can isolate the contribution of the single-proton dynamics to the effective lattice 
potential.
In particular, we considered two types of embedding: (i) the situation where the proton feels a potential energy surface (PES) determined by a classical environment, i.e., the other ions are fixed at their static equilibrium positions, and (ii) the quantum embedding where the surrounding ions are treated as quantum particles, producing a PES which includes quantum fluctuations (see SM \cite{SM}). 
These two PES are shown in Fig.~\ref{fig:fig4}(a) and (b), and their difference is shown in Fig.~\ref{fig:fig4}(c). The 
collective quantum fluctuations 
smooth out the potential, leading to a more cylindrical PES around the $x$ axis. 
In Fig.~\ref{fig:fig4}(e), we report  in dashed lines the minimum energy of the PES at each $d_{OO}$ for the two embeddings, defining 
$V_{OO}$ in the two cases. 
To examine whether local quantum effects
introduce anharmonicity into $V_{OO}$, we calculated the proton ZPE by solving the effective three-dimensional Schr\"odinger equation in the evaluated PES \cite{DiCataldo2024, Renaud2026} for each $d_{OO}$ (see SM \cite{SM}). The ZPE evolution as a function of $d_{OO}$ is shown in Fig.~\ref{fig:fig4}(d) and the 
ZPE-corrected
$V_{OO}$ is plotted in full lines in Fig.~\ref{fig:fig4}(e).
In all cases, the shape of $V_{OO}$ retains its harmonic character. Indeed, the ZPE grows linearly with volume and does not introduce measurable anharmonicity at this energy scale. 
The $V_{OO}$ minimum determines the equilibrium $d_{OO}$ and, hence, the corresponding density at 0 K for each model, as plotted in Fig.~\ref{fig:fig2}. As expected, the density for a classical proton in the classical embedding falls exactly on the density extrapolated to 0 K from the MD simulations, because in both approaches all ions are classical. 
Remarkably, the anomalous increase of the proton ZPE with volume shifts the equilibrium $d_{OO}$ towards lower values resulting in higher equilibrium densities (filled blue point in Fig.~\ref{fig:fig2}). Therefore, treating only the embedded proton as a quantum particle is not sufficient to explain the anomalous behavior of the density. Accurately reproducing the experimental and PIMD densities requires accounting for the quantum nature of both the proton and its surrounding environment (open red point in Fig.~\ref{fig:fig2}). 
Notably, although our embedding approach neglects the explicit coupling between the embedded particle and its surroundings, it
successfully captures the essential physics of the system:
the NTE is a collective quantum phenomenon.



In addition to the importance of collective quantum effects,
the effective potential $V_{OO}$ obtained by our model brings about another 
feature. Indeed, its quadratic shape 
rules out classical anharmonicity as the primary driver of the NTE \cite{Min2025}.
This 
rationalizes the QHA results, which reproduce 
a large part of the 
quantum renormalisation of volume obtained from fully anharmonic PIMD.
In fact, QHA calculations by Pamuk \emph{et al.} \cite{Pamuk2012} recover about 70\% of the relative volume change induced by 
NQE
at 0 K, obtained by extrapolating 
our classical MD and PIMD calculations. 

\begin{figure}[H]
    \centering
    \includegraphics[width=1\linewidth]{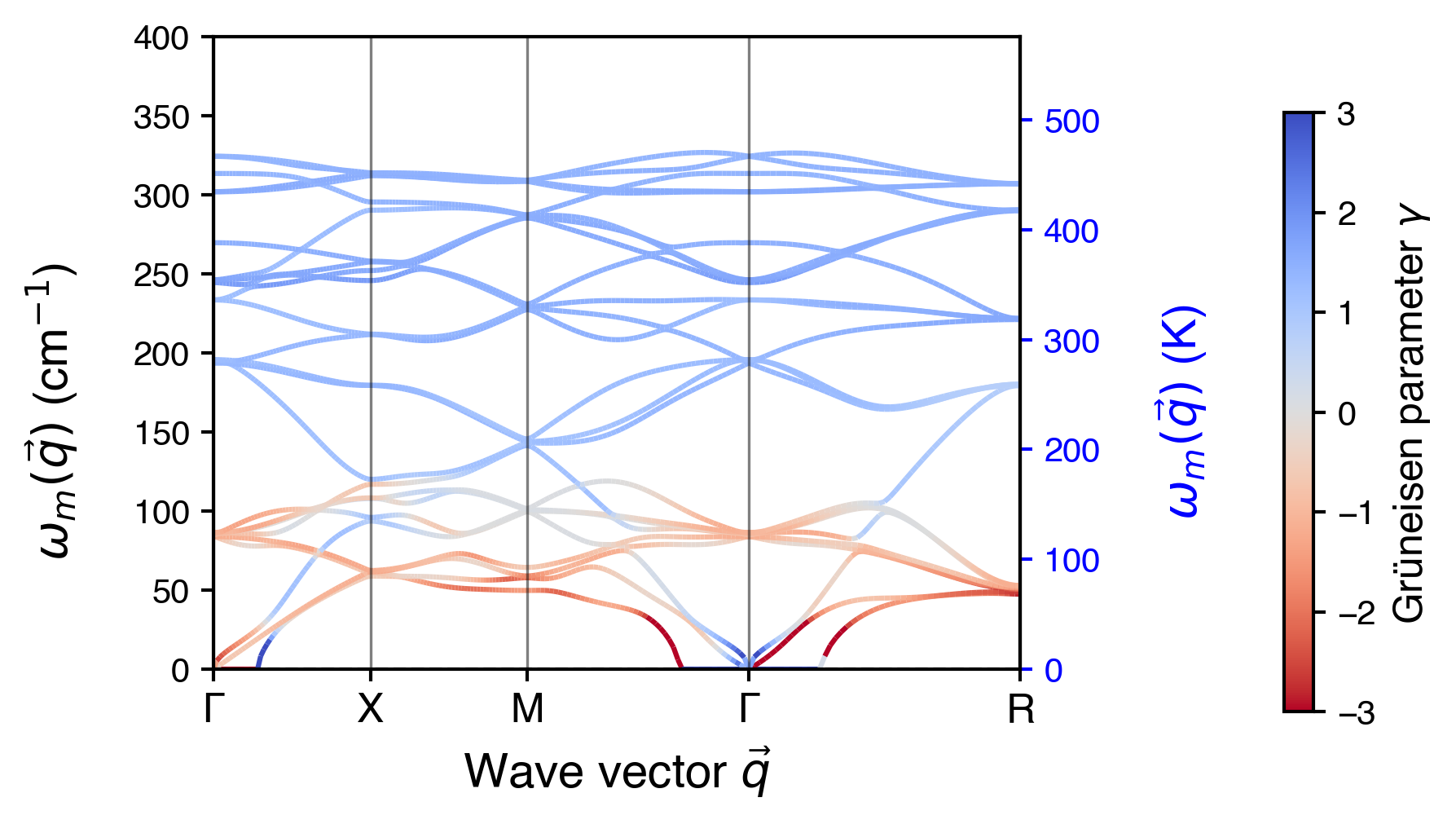}
    \caption{Phonon dispersion  of the low-lying modes in the conventional cubic cell at the classical equilibrium volume ($\omega_{m}(\vec{q},V_\textrm{eq})$) obtained from 
    frozen-phonon
    calculations with MB-pol. The phonon-band colors encode the Grüneisen parameter, with positive (negative) values in blue (red). The imaginary modes near the $\Gamma$ point are 
    artifacts 
    of the $\vec{q}$-mesh interpolation. Long-wavelength analytic corrections have not been included in the dynamical matrix. 
    }
    \label{fig:fig5}
\end{figure}

As far as NTE is concerned, the comparison between QHA and PIMD estimates 
is even more favorable, with almost the same density slope and maximum temperature.
The overall discrepancies between QHA and PIMD 
could either come from the approximation behind QHA, which 
neglects anharmonic effects, or may arise from differences between the PES used in the two studies.
The NTE is quantified
by the volumetric thermal expansion coefficient $\alpha_V = \frac{1}{B_T}\frac{\partial^2 F}{\partial T \partial V}$, with $B_T$ the isothermal bulk modulus and $F$ the free energy, which in QHA takes the following form for the vibrational part\cite{Born1955}:
\begin{equation}
    F^{vib}_{QHA}=\sum_{m,\vec{q}} \left( \frac{\hbar \omega_{m}(\vec{q}, V)}{2} + k_B T
    \ln\left(1-e^{-\frac{\hbar \omega_{m}(\vec{q}, V)}{k_B T}}  \right) \right).
    \label{eq:entropy}
\end{equation}
Negative contributions to $\alpha_V$ come from quantized vibrational modes $\omega_{m}(\vec{q},V)$ with a negative Grüneisen parameter $\gamma \equiv - V \partial_{V} \ln(\omega_{m}(\vec{q},V)) $, i.e., those whose frequency increases with volume $V$ (see SM\cite{SM}).
To pinpoint the specific modes driving the contractive 
$F^{vib}_{QHA}$ 
response as a function of temperature $T$, we mapped mode-resolved Grüneisen parameters onto the harmonic phonon dispersion, computed via finite differences using the MB-pol interatomic potential in a $3 \times 3 \times 3$ supercell of the I$_c$ conventional cubic cell (Fig.~\ref{fig:fig5}).
Analysis of the dispersion reveals that intramolecular stretching ($\sim 3200-3400 ~ \textrm{cm}^{-1}$) and bending ($\sim 1600-1800 ~\textrm{cm}^{-1}$) modes exhibit only marginally negative $\gamma$ (see SM \cite{SM}).
Instead, the dominant contractive contributions—characterized by the largest negative values—originate from the transverse acoustic (TA) branches and the lowest-lying transverse optical (TO) ones, particularly along $\Gamma-\textrm{R}$, that is, a $\langle111 \rangle $ direction of the cubic tetrahedral network, i.e. one of the four equivalent O--H$\cdots$O bond directions.
This is consistent with the structural anisotropy observed in Fig.~\ref{fig:fig3}, which associates the 
transverse-over-longitudinal spread of quantum proton fluctuations 
with the NTE.
By computing PIMD phonon frequencies \cite{Morresi2021}, we verified that these transverse vibrations, despite their strong hydrogen character, remain close to the harmonic limit, whereas
appreciable quantum-anharmonic renormalization appears at higher frequencies (see SM \cite{SM}).

The macroscopic NTE evolution 
is thus dictated by how 
$F^{vib}_{QHA}$
in Eq.~\ref{eq:entropy} changes according to these branches, sequentially populated 
as the temperature rises. At the lowest temperatures, the contractive response is governed by the TA branches, whose Bose-Einstein populations grow first (see SM \cite{SM}). These correspond to the hydrogen-bond bending modes previously identified in ice I$_h$ via pressure-dependent inelastic neutron scattering \cite{Strssle2004}. As $T$ increases within the NTE regime, an additional contractive contribution comes from the low-lying TO branches, as shown in SM. These modes have characteristic energies corresponding to $T \approx$ 60--80~K, and involve collective transverse translations of H$_2$O molecules \cite{Li1993}. Above $\sim$70 K, the thermal excitation of higher-frequency modes with positive $\gamma$
overtakes the contractive $F^{vib}_{QHA}$ term, marking the transition to standard positive thermal expansion.

This statistical $T$-dependence fundamentally explains why classical MD fails to reproduce the NTE despite using the same MB-pol potential. In classical MD, the modes population follows equipartition rather than Bose-Einstein statistics. Consequently, the selective low-temperature weighting of the negative-$\gamma$ branches is lost, and the positive-$\gamma$ modes dominate the net response. Therefore, negative Grüneisen parameters are a necessary but not sufficient condition to explain the emergence of NTE; the quantum-statistical weighting of the different phonon branches is essential.

The direct comparison between I$_c$ and I$_h$ provides a uniquely controlled benchmark for atomistic models of ice: their closely similar density maxima show that NTE in ice I is not tied to a particular stacking sequence, but is an intrinsic response of the open tetrahedral hydrogen-bond network. Our results connect three levels of description: the anisotropic quantum distribution of the proton, the low-frequency transverse network modes with negative Grüneisen parameters, and the quantum-statistical weighting of these modes in the vibrational free energy. The NTE is therefore not a local zero-point correction to an individual O--H bond, but a collective nuclear quantum response of the hydrogen-bond network. 
This framework also suggests why comparable cryogenic NTE is not generally observed in denser ice polymorphs, where compression stiffens the low-frequency network modes and modifies proton delocalisation. Stacking-disorder-free ice I$_c$ not only  provides direct evidence for the network-driven origin of NTE in water, but is also a minimal, chemically transparent reference for quantum-driven thermodynamic anomalies in other open tetrahedral networks.

\textit{Acknowledgments} - We acknowledge the Institut Laue-Langevin for providing beamtime under the proposal number 5-25-295. 
L.E.B., T.P., and R.G.\ acknowledge financial support from the Swiss National Fund (FNS) under Grant No.~212889.
L.R.\ and A.M.S.\ acknowledge GENCI for providing computational resources on the CINES Adastra and IDRIS Jean-Zay supercomputing clusters under project numbers 2025-A0180901387 and 2026-A0200901387.
L.R.\ and M.C.\ acknowledge GENCI for providing computational resources on the IDRIS Jean-Zay supercomputing clusters under project number 2024-A0170906493 and 2025-A0190906493.
M.C.\ thanks the European High Performance Computing Joint Undertaking (JU) for partial support through the ``EU-Japan Alliance in HPC'' HANAMI project (Hpc AlliaNce for Applications and supercoMputing Innovation: the Europe--Japan collaboration).
L.R. acknowledges Marco Cherubini for fruitful discussions about anharmonic phonons. L.R. and A.M.S. acknowledge Camille Jolette for her assistance with the early-stage harmonic phonon calculations during her internship at LPENS.

\nocite{litman2024ipi, Babin2014, Babin2013, Medders2014, Riera2023, Paesani2016, Reddy2016, Matsumoto2017bk, Matsumoto2024, Uhl2016, Ceriotti2009, Ceriotti2010, phonopy-phono3py-JPSJ, phonopy-phono3py-JPCM, Morresi2021, Morresi2022, Giannozzi_2017, Giannozzi_2009, Renaud2026, Monacelli2025, Monacelli_2021}

\bibliography{pnas-sample}

\end{document}


\setcounter{figure}{0}
\renewcommand{\thefigure}{S\arabic{figure}}
\title{Supplemental Material: {Negative Thermal Expansion in Cubic Ice: A Collective Quantum Effect of the hydrogen-bond network}}

\author{Loan Renaud}
\email{loan.renaud@ens.fr}
\affiliation{Laboratoire de physique de L'École normale supérieure de Paris, CNRS, ENS \& Université PSL, Sorbonne Université, Université de Paris, 75005 Paris, France}
\author{Tomasz Poreba}\affiliation{Laboratory for Quantum Magnetism, Institute of Physics, \'{E}cole Polytechnique F\'{e}d\'{e}erale de Lausanne, CH-1015 Lausanne, Switzerland}
\author{Richard Gaal}\affiliation{Laboratory for Quantum Magnetism, Institute of Physics, \'{E}cole Polytechnique F\'{e}d\'{e}erale de Lausanne, CH-1015 Lausanne, Switzerland}
\author{A. Marco Saitta}\affiliation{Laboratoire de physique de L'École normale supérieure de Paris, CNRS, ENS \& Université PSL, Sorbonne Université, Université de Paris, 75005 Paris, France}
\affiliation{Institut Universitaire de France (IUF)}
\author{Michele Casula}\affiliation{Institut de Min\'{e}ralogie, de Physique des Mat\'{e}riaux et de Cosmochimie (IMPMC), Sorbonne Universit\'{e}, CNRS UMR 7590, MNHN, 4, place Jussieu, Paris, France}
\author{Livia Eleonora Bove}\affiliation{Laboratory for Quantum Magnetism, Institute of Physics, \'{E}cole Polytechnique F\'{e}d\'{e}erale de Lausanne, CH-1015 Lausanne, Switzerland}\affiliation{Institut de Min\'{e}ralogie, de Physique des Mat\'{e}riaux et de Cosmochimie (IMPMC), Sorbonne Universit\'{e}, CNRS UMR 7590, MNHN, 4, place Jussieu, Paris, France}
\affiliation{Dipartimento di Fisica, Sapienza Universit\`a di Roma, Piazzale Aldo Moro 5, 00185 Roma, Italy}

\date{\today}
\maketitle

\section{Neutron-diffraction methods}

In this work, we synthetize the sample following the procedure of Komatsu et al. to produce pure cubic ice by controlled degassing of C2 hydrogen hydrate \cite{Komatsu2020}. First, an approximately equimolar mixture of sII hydrogen hydrate and MgD$_2$ approximately 30 mm$^3$ (oppure about (oppure about 30 mm$^3$)  were loaded together into a TiZr gasket in a liquid nitrogen bath. The gasket was closed, pressurized to ~0.3 GPa in a Paris-Edinburgh press, and heated to 85 $^\circ C$ for 2 hours to produce excess deuterium gas and brucite as a byproduct. The mixture was cooled down to room temperature and compressed inside the TiZr gasket to 3.9 GPa, and subsequently cooled to 70 K in Paris-Edinburgh press, to produce C2 hydrogen hydrate. The press was then opened while kept immersed in a liquid-nitrogen bath. The extracted gasket was placed in the vanadium can kept at 70 K, and transferred to a  cryostat pre-cooled to 50 K. Neutron diffraction data ($\lambda$ = 1.54 \AA) from the sample were collected on the D20 diffractometer at the ILL. The resulting diffractograms were refined using TOPAS software \cite{Coelho2018} using either the Le Bail method (for lattice parameter extraction) or Rietveld refinement to extract the anisotropic displacement parameters for D atoms at 50 K. Diffractograms before and after decompression show marked differences. Firstly, the 111 reflection of C2, isostructural to cubic ice, is largely suppressed due to the presence of disordered D$_2$ molecules at the  48f Wyckoff site (x, 1/8, 1/8) in a cubic \textit{Fd$\bar{3}$m} structure. Upon decompression and recooling to 50 K this reflection dominates the diffractogram. The refined D$_2$ occupancy at this site is 0.00(4), confirming that the sample is hydrogen-free. Along with cubic ice, remnant ice VIII, brucite, and traces of unreacted MgD$_2$ are present, while cubic ice itself shows neither stack-disorder nor admixtures of hexagonal ice. The pure I$_c$ was then subjected to a controlled heating program (circa 0.5 K/min) and diffraction data was collected every 15 K up to 200 K. Temperature was measured by using a K-type thermocouple attached in the vicinity of the one of the anvils. Each measurement lasted for about 1 hour. Temperature equilibration was reached typically within 15 minutes. Heating above 200 K resulted in the formation of a large fraction of stack-disordered ice and eventually transition to hexagonal ice, which precluded a reliable determination of the unit-cell parameters. determination. The unit-cell parameters of cubic ice as a function of temperature were extracted using Le Bail method \cite{LeBail_2005}.

\begin{table}[htpb]
    \centering
    \caption{Temperature dependence of the lattice parameter $a$ and the associated error $da$}
    \label{tab:latparam}
    \begin{tabular}{lccccccccccc}
        \hline
        T [K] & 50 & 65 & 80 & 95 & 110 & 125 & 140 & 155 & 170 & 185 & 200 \\
        \hline
        a [\AA{}] & 6.3551 & 6.3543 & 6.3544 & 6.3553 & 6.3559 & 6.3571 & 6.3581 & 6.3598 & 6.3628 & 6.3684 & 6.3724 \\
        da [\AA{}] & 0.0008 & 0.0009 & 0.0009 & 0.0009 & 0.0010 & 0.0010 & 0.0010 & 0.0008 & 0.0007 & 0.0007 & 0.0007 \\
        \hline
    \end{tabular}
\end{table}

\section{Input file used for anisotropic displacement parameter structure refinement of I$_c$  in TOPAS software \cite{Coelho2018}}
Input file (INP) containing all the observed phases, refined anisotropic displacement parameter (ADPs) for D atoms and refinement details are available in the Zenodo repository \cite{poreba_2026_21676175}.

\section{Crystallographic information file generated from experimental neutron diffraction data collected at 50 K}

Crystallographic information file (CIF) file contains an experimental I$_c$ structural model (data collected at 50 K) with D atoms refined anisotropically. It is available in the Zenodo repository \cite{poreba_2026_21676175}.

\section{Density calculation from molecular dynamics}

The density at atmospheric pressure ($P=1$ atm) was calculated as a function of temperature using Path-Integral Molecular Dynamics (PIMD) simulations performed with the i-PI universal force engine \cite{litman2024ipi}. The energy and forces have been obtained using the MB-pol data-driven many-body potential included in the MBX library \cite{Babin2014, Babin2013, Medders2014, Riera2023, Paesani2016,Reddy2016}.

The structures have been generated with GenIce \cite{Matsumoto2017bk, Matsumoto2024} using a $3\times3\times3$ replicated cell. They contain 648 atoms for cubic ice and 1296 atoms for hexagonal ice.

For each temperature,  short equilibration runs ($\sim 20$ ps) are first performed in the canonical NVT ensemble using a Langevin thermostat with a time step $\Delta t=0.25\text{ fs}$ and a relaxation time $\tau = 1 \text{ ps}$. From the output of that equilibration, a second equilibration ($\sim 125$ ps) is performed in the isothermal isobaric NPT ensemble with the same thermostat and an isotropic Langevin barostat with relaxation time $\tau = 10 \text{ ps}$. After that, PIMD simulations are performed using the PIGLET \cite{Uhl2016} colored-noise thermostat with $\omega_{max} \ge 3700 \text{ cm}^{-1}$ and 8 additional fictitious momenta. For the barostat, we use an optimal sampling colored-noise thermostat with 4 additional fictitious momenta. The parameters for the generalized Langevin equations have been generated by GLE4MD \cite{Ceriotti2009, Ceriotti2010}.

The number of replicas (beads) $P$ and timestep have been chosen to converge both the density of the system and the kinetic estimators. P was chosen such that the product $T\times P$ stays constant. We used  $P=48$ at $T=100$ K, five times fewer than the number of beads required with the local Path Integral Langevin Thermostat (PILE-L) to reach the same convergence. The production runs lasted approximately $100$ ps.

\section{Mass rescaling}

In Figure 2, we directly compared our H$_2$O simulation results with experimental D$_2$O measurements using a mass-based scaling factor $\dfrac{M_{H_2O}}{M_{D_2O}}$. To validate this procedure, we applied the same scaling to compare D$_2$O and H$_2$O PIMD simulations.\\ 

\begin{figure}[h!]
    \centering
    \includegraphics[width=0.6\linewidth]{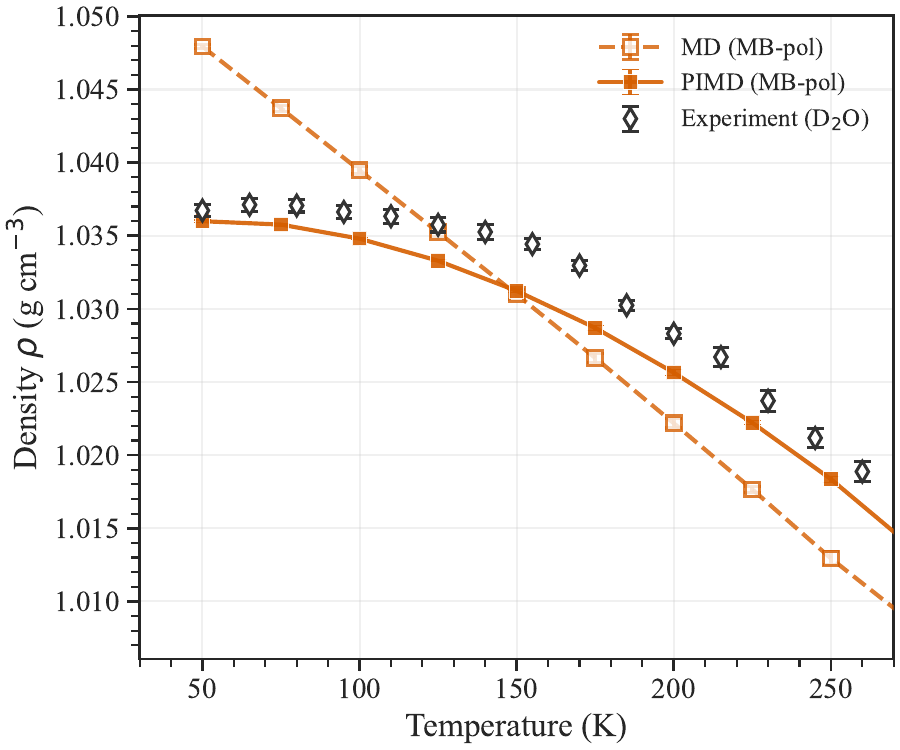}
    \caption{D$_2$O densities obtained from MD and PIMD simulations together with the original neutron-diffraction data.}
    \label{fig:densityD2O}
\end{figure}

Figure \ref{fig:densityD2O} displays the molecular dynamics (MD) and PIMD simulations of deuterated ice—performed under the previously described conditions—plotted against the unscaled experimental data. The PIMD predictions show close agreement with the experimental measurements.\\

\begin{figure}[h!]
    \centering
    \includegraphics[width=0.6\linewidth]{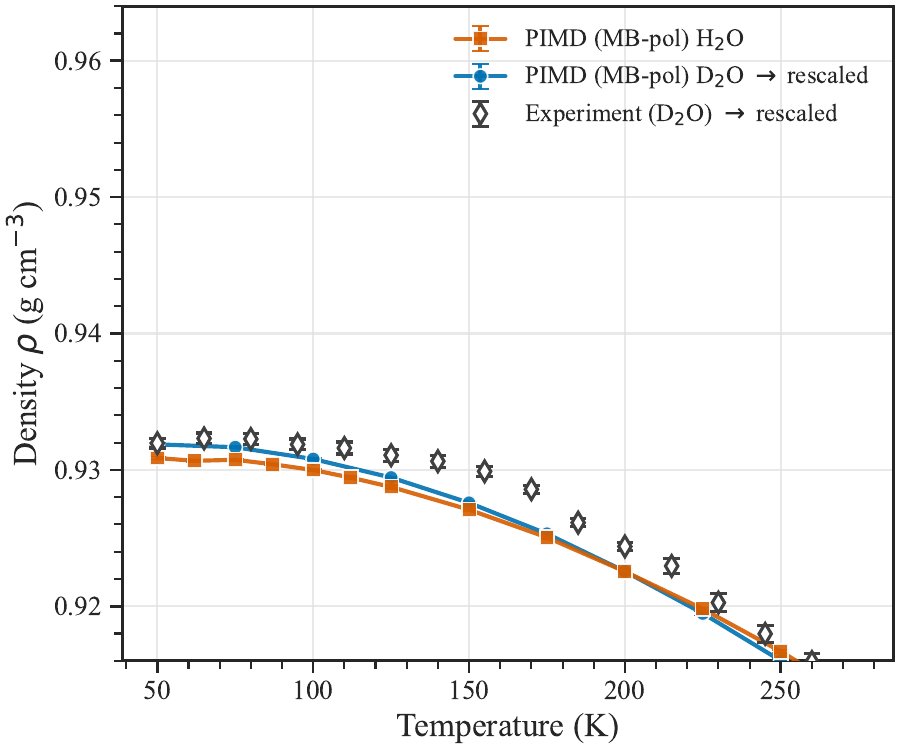}
    \caption{Comparison of the H$_2$O PIMD density with mass-rescaled D$_2$O PIMD and experimental densities.}
    \label{fig:massrescaled}
\end{figure}

Finally, Figure~\ref{fig:massrescaled} illustrates that rescaling the D$_2$O experimental and PIMD densities by the mass factor causes them to collapse onto the H$_2$O PIMD data. The deviation between the rescaled D$_2$O and H$_2$O curves falls well within the experimental uncertainty region, supporting the mass-rescaling procedure used in Figure 2 of the main text. 

\section{Zero-point energy calculations}
\begin{figure}[h!]
    \centering
    \includegraphics[width=0.5\linewidth]{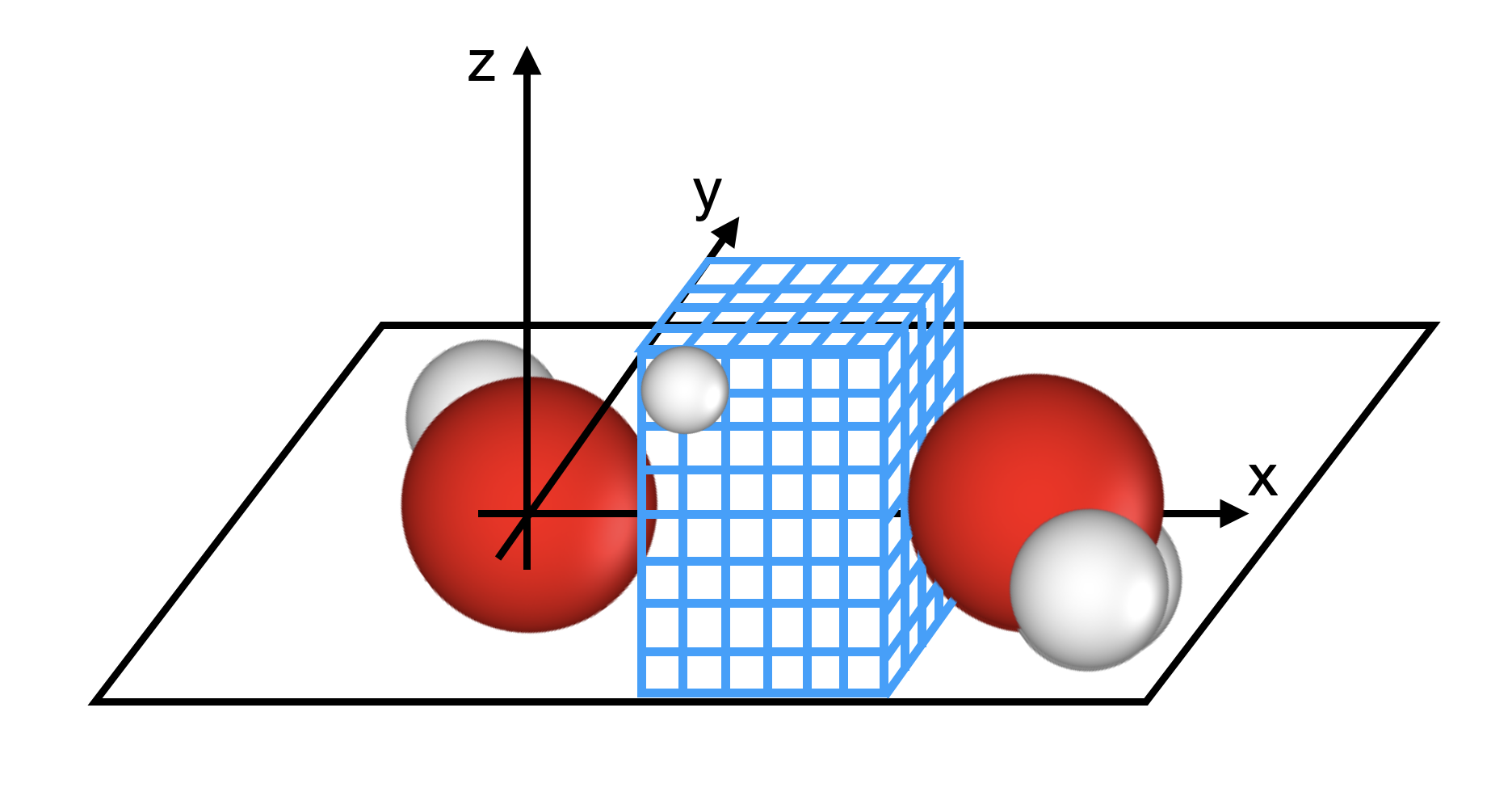}
    \caption{Schematic representation of the grid used to scan the potential energy surface (PES) experienced by the proton involved in the $\text{O}-\text{H} \cdots \text{O}$ bond. The potential of the particle is evaluated at each grid point (see the next Section for details).}
    \label{fig:S1}
\end{figure}

The effect of the zero-point energy (ZPE) on the classical effective O--O interatomic potential $V_{OO}$ as a function of O--O distance $d_{OO}$ is evaluated by scanning the Potential Energy Surface (PES) felt by a proton and generated by the surrounding water molecules, treated either as classical or as quantum ions, on a dense grid (see Figure~\ref{fig:S1}) and solving the Schrödinger equation with this PES for different cell volumes $V$, or equivalently different $d_{OO}$ distances, since $d_{OO}=\frac{\sqrt{3}}{4}V^{\frac{1}{3}}$ for cubic ice.

\subsection{Potential energy surface scan}

\begin{figure}[h]
    \centering
    \includegraphics[width=1\linewidth]{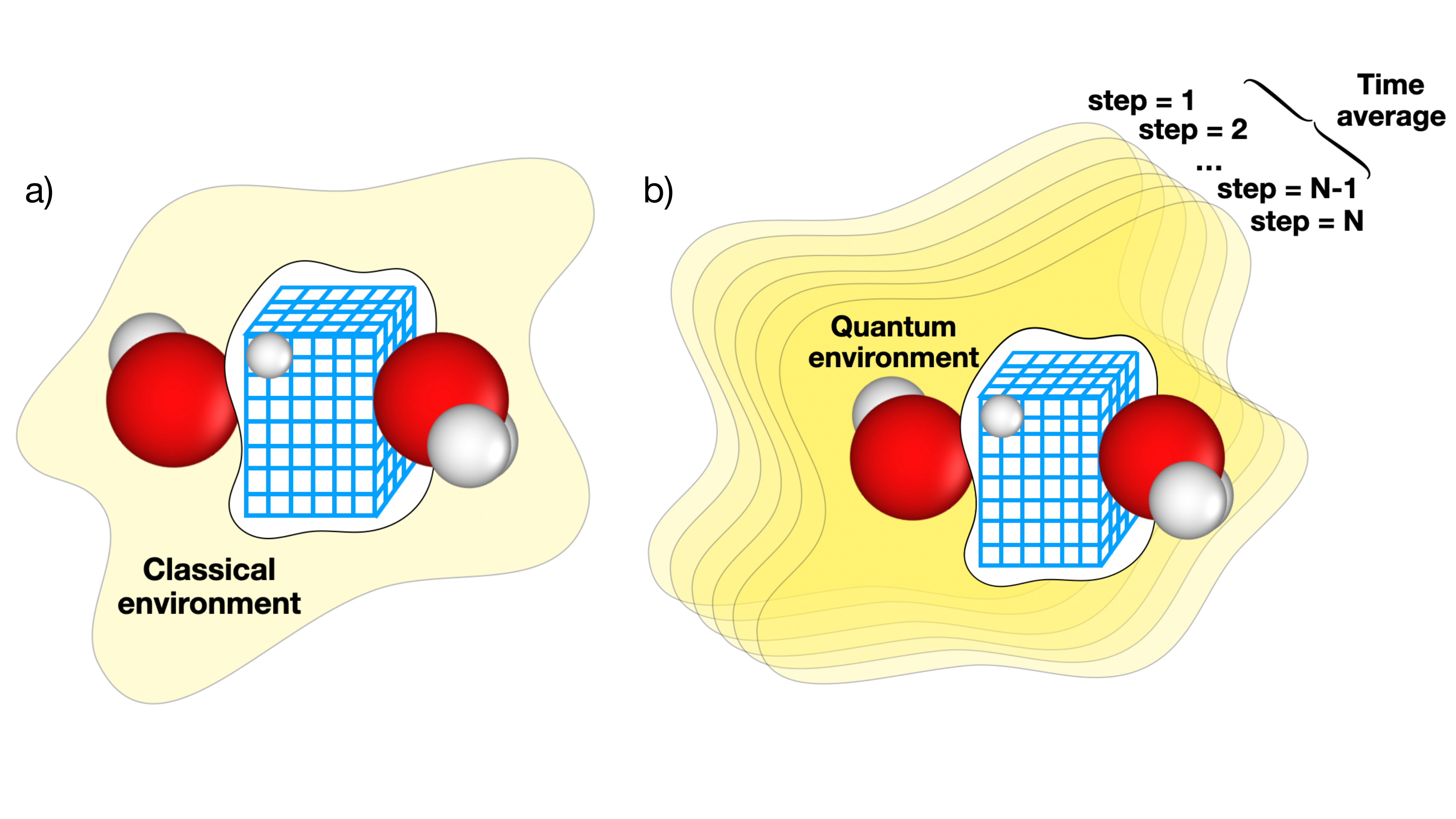}
    \caption{(a) Evaluation of the PES derived from relaxation calculations mapped onto a 3D grid defining the proton position between the two oxygen atoms.
    (b) Evaluation of the PES obtained by time-averaging NVT-PIMD trajectories sampled at each grid point, with the $\text{O}-\text{H} \cdots \text{O}$ coordinates constrained.}
    \label{fig:S2}
\end{figure}

\textbf{Classical embedding - } The classical embedding is defined such that at each proton position of the three dimensional (3D) grid, a geometry optimization has been performed on the H atomic positions of the $2\times2\times2$ supercell using MB-pol and the resulting energy is used to build the PES (oxygen atoms are kept fixed). 
This approach can be seen as the embedding of a quantum particle (the hydrogen atom of interest) in an environment made of classical particles. This approach is depicted in Figure~\ref{fig:S2}(a).\\

\textbf{Quantum embedding - }
In order to assess the effect of quantum and thermal delocalization of the surrounding particles on the proton, we used a second approach to evaluate the quantum embedding version of the PES. A set of PIMD simulations in the canonical ensemble is performed for different cell volumes fixing the positions of 3 atoms forming the hydrogen bond $\text{O}-\text{H} \cdots \text{O}$, see  Figure~\ref{fig:S2}(b). While the O-O bond length ($d_{OO}$) is kept fixed for the whole set of simulations (to match the desired volume value), each trajectory is evaluated for a different position of the H atom on a 3D grid, labelled as $(x_{H,i}, y_{H,i}, z_{H,i})$ with $i \in \{1,..., N\}$ where $N$ is the number of grid points. By integrating the energy of the surrounding particles for the fixed set $(d_{OO}; x_{H,i}, y_{H,i}, z_{H,i})$ over each trajectory (as schematized on  Figure~\ref{fig:S2}(b)), one can build an estimation of the PES on the 3D grid for each volume. The trajectories have been calculated at $T=80$ K using the PIGLET colored-noise thermostat and after equilibration, around 300 frames have been used to estimate the value of each PES grid point. By doing so, we derive the proton quantum embedding.

\subsection{Resolution of the Schrödinger equation for a single proton in the potential energy surface}
After interpolating the binned PES, an estimation of the ZPE associated to the PES corresponding to each $d_{OO}$ value is obtained via two methods.\\

\textbf{Method 1 - }The first method consists in reducing the 3D PES to a numerically tractable two dimensional (2D) potential by assuming that the PES is cylindrically symmetric around the O--O bond (i.e. supposing it is invariant under a rotation around the $x$ axis). The Schrödinger equation for the H atom in the 2D cylindrical PES $V(\rho, z)$ can then be solved numerically.\\

\textbf{Method 2 - }The second method yields the ZPE after averaging over the energy of an H atom simulated via PIMD for the given PES at a low temperature ($T \approx 25 K$) and substracting the energy at the bottom of the potential well.\\

\begin{figure}[h!]
    \centering
    \includegraphics[width=0.5\linewidth]{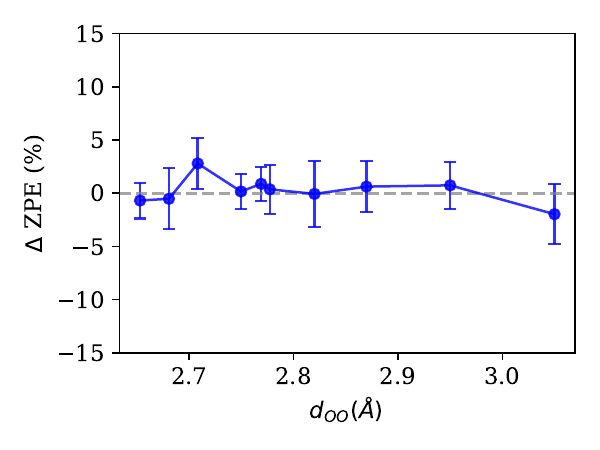}
    \caption{Relative difference between the ZPE evaluated using method 1 (numerical resolution of the Schrödinger equation) and method 2 (PIMD trajectory sampling the proton in the PES) as a function of the O--O distance $d_{OO}$.}
    \label{fig:S3}
\end{figure}

Figure~\ref{fig:S3} illustrates the good agreement between the two methods. The relative difference between the two independent ZPE calculations remains below 5\%, in statistical agreement within the uncertainty range.

\section{MODE-RESOLVED GRÜNEISEN PARAMETERS}
Harmonic phonons were computed via the finite difference method implemented in the i-PI engine with the MB-pol potential. Atomic displacements of 0.01 \AA{} were applied to a $3 \times 3 \times 3$ supercell at three distinct volumes: the equilibrium volume ($V_0$), a 3\% compression ($V_0 - 3\% V_0$), and a 2\% expansion ($V_0 + 2\% V_0$). The Phonopy package \cite{phonopy-phono3py-JPSJ, phonopy-phono3py-JPCM} was employed to evaluate the phonon dispersion along the $\Gamma-X-M-\Gamma-R$ path in the Brillouin zone and to calculate the phonon density of states (DOS) using a $25 \times 25 \times 25$ uniform $\mathbf{q}$-point mesh.
The complete phonon dispersion, including both low- and high-frequency branches, is shown in Fig. \ref{fig:dos}(a), and the corresponding DOS is reported in Fig. \ref{fig:dos}(b).
\section{Quantum anharmonic phonons}
To capture full quantum anharmonic effects, we evaluated the phonons using the framework introduced in Refs. \cite{Morresi2021, Morresi2022}, which relies on zero-time Kubo-transformed correlation functions computed from PIMD samples. We performed a PIMD simulation on a $3 \times 3 \times 3$ proton-ordered supercell in the canonical (NVT) ensemble at 75 K. The system was coupled to a local PILE thermostat with a time step of $\Delta t = 0.25$ fs and a relaxation time of $\tau = 100$ fs.
Within this methodology, we obtained two independent estimates of the dynamical matrices: one derived from displacement-displacement ($uu$) correlators and the other from force-force ($ff$) correlators. The resulting dynamical matrices were symmetrized, and the acoustic sum rule imposed using CellConstructor package \cite{Monacelli_2021}. Finally, real-space interatomic force constants were extracted using the \texttt{q2r.x} routine from the Quantum ESPRESSO suite \cite{Giannozzi_2017, Giannozzi_2009}, and the corresponding phonon DOS was interpolated on a $25 \times 25 \times 25$ uniform $\mathbf{q}$-mesh using \texttt{matdyn.x}.

\begin{figure}[h]
    \includegraphics[width=0.49\linewidth]{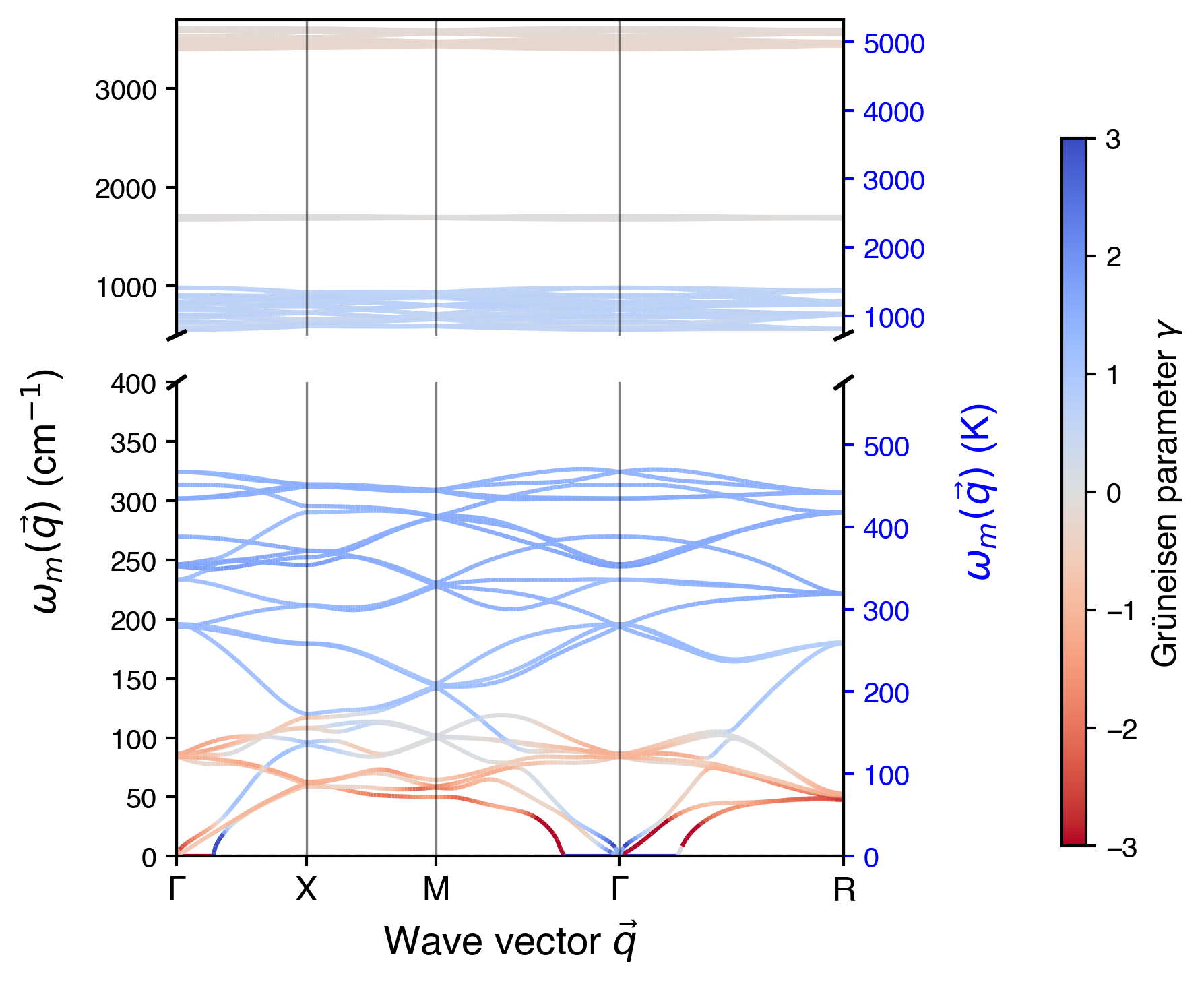}
    \includegraphics[width=0.49\linewidth]{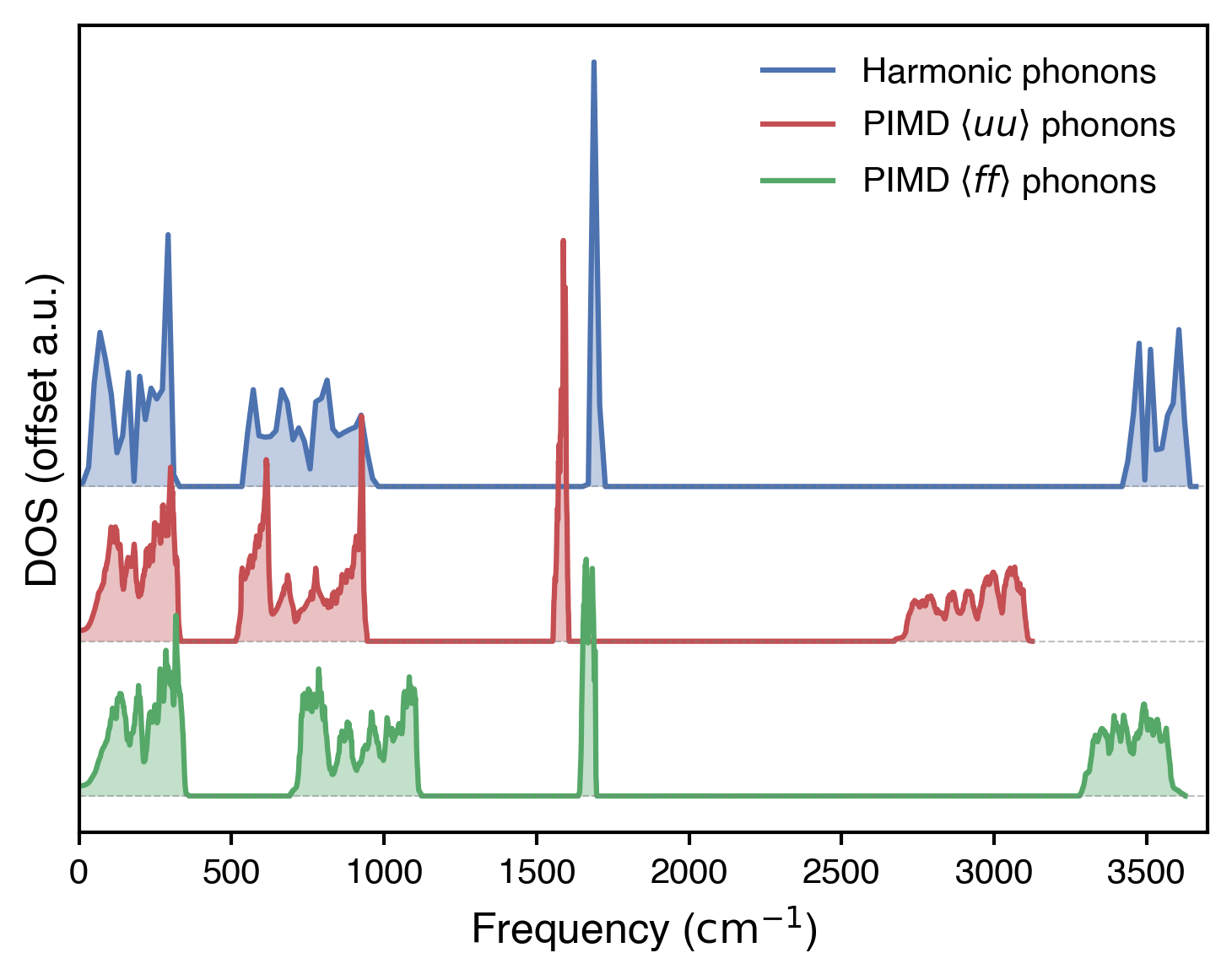}
    \caption{(a) Phonon dispersion in the conventional cubic cell at the classical equilibrium volume ($\omega_{m}(\vec{q},V_\textrm{eq})$) obtained from frozen-phonon calculations with MB-pol. The phonon-bands color encodes the Grüneisen parameter, with positive (negative) values in blue (red). The imaginary modes near the $\Gamma$ point are artifacts of the $\vec{q}$-mesh interpolation. Long-wavelength corrections have not been included in the dynamical matrix.  (b) Comparison between harmonic phonon DOS (finite differences) and quantum anharmonic DOS extracted from $⟨uu⟩$ and $⟨ff⟩$ correlators via PIMD.}
    \label{fig:dos}
\end{figure}

We display our two quantum anharmonic evaluations of the DOS in Fig. \ref{fig:dos} alongside with the DOS obtained from harmonic phonons. Interestingly, while both $⟨uu⟩$ and $⟨ff⟩$ correlators are extracted from the same anharmonic PIMD trajectories, they provide different effective phonon frequencies and therefore different DOS. 

The force-force correlator reflects the thermally averaged local curvature of the potential energy surface, maintaining a more harmonic-like character. In contrast, the displacement-displacement correlator is directly related to the definition of the phonon Green's function and captures the full amplitude of the atomic displacement, providing a more robust measure of global anharmonic effects such as phonon softening. The two phonon estimators coincide in the harmonic limit \cite{Morresi2021}.
The libration ($\sim 600-1000 \textrm{ cm}^{-1}$), bending ($\sim 1700 \textrm{ cm}^{-1}$) and stretching ($\sim 3400-3600 \textrm{ cm}^{-1}$) modes are largely affected by quantum anharmonicity, quantified by the shift between the $⟨uu⟩$ and $⟨ff⟩$ bands and the one with respect to the full harmonic case. 
The softening of the stretching mode computed with $⟨uu⟩$ is about $\sim 500 \textrm{ cm}^{-1}$, the same amount as the one we already observed in the cubic water skeleton of C2 hydrogen hydrate for one of our previous studies \cite{Renaud2026} and compatible with SSCHA data published in \cite{Monacelli2025}.
Conversely, the low-frequency band ($\sim 0-350 \textrm{cm}^{-1}$) shows no significant anharmonic renormalisation. This explains why the Quasi-Harmonic Approximation (QHA) reproduces well the Negative Thermal Expansion (NTE) behavior since, as demonstrated in our work, these low-frequency modes are the ones involved in the NTE. 

\section{Mode resolved thermal expansion coefficient}

We evaluated the mode resolved thermal expansion coefficient from our computed Grüneisen parameters.
Within the QHA framework, the volumetric thermal expansion coefficient $\alpha_V(T)$ is driven by the weighted sum of the modal heat capacities $c_{v}(q, \nu, T)$:
$$\alpha_V(T) = \frac{1}{V_{eq} B} \sum_{\vec{q},m} \gamma_{\vec{q},m} \, c_{v}(\vec{q}, m, T)$$
where $B$ is the isothermal bulk modulus and $c_{v}(q, \nu, T)$ is the quantum harmonic heat capacity for a mode of frequency $\omega_{\vec{q},m}$ at temperature $T$:
$$c_{v}(\vec{q},m, T) = k_B \left( \frac{\hbar \omega_{\vec{q},m}}{k_B T} \right)^2 \frac{e^{\frac{\hbar \omega_{\vec{q},m}}{k_B T}}}{\left( e^{\frac{\hbar \omega_{\vec{q},m}}{k_B T}} - 1 \right)^2}$$
with $k_B$ the Boltzmann constant, $\hbar$ the reduced Planck constant and  $\omega_{\vec{q},m}$ the phonon frequency for wavevector $q$ and branch $m$.

\begin{figure}[h!]
    \centering
    \includegraphics[width=0.7\linewidth]{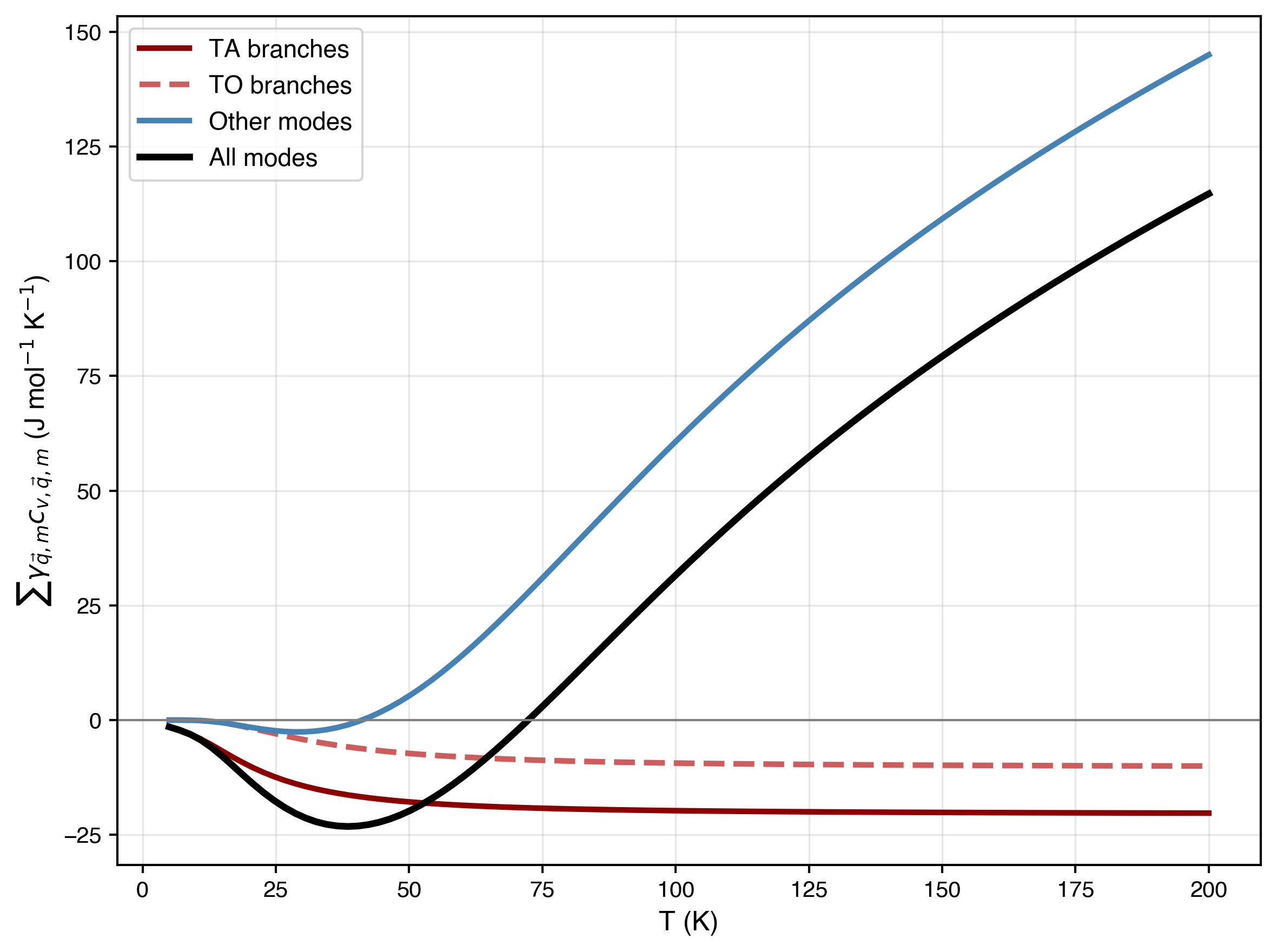}
    \caption{Mode resolved contributions to the thermal expansion coefficient as a function of temperature.}
    \label{fig:alpha}
\end{figure}

To demonstrate which modes are responsible for the NTE, we partitioned this thermodynamic sum into specific band contributions (see Figure \ref{fig:alpha}). Specifically, we isolated the thermodynamic weight, $\sum \gamma_{\vec{q},m} c_v(\vec{q}, m, T)$, of the transverse acoustic (TA) and low-lying transverse optic (TO) modes, separating them from the rest of the higher-frequency spectrum.
As shown in Figure \ref{fig:alpha}, the mode-resolved analysis reveals a clear temperature-dependent progression that dictates the macroscopic volumetric behavior:
\begin{itemize}
    \item at low temperatures ($T < 30$ K): the TA modes, being the lowest energy excitations, are the first to be thermally populated. Because these modes exhibit strongly negative Grüneisen parameters, they entirely dominate the thermodynamic sum, initiating the NTE;
    \item at intermediate temperatures ($30$ K $< T < 60$ K): as the temperature increases, the low-lying TO modes become thermally accessible. These modes also possess negative Grüneisen parameters and add a secondary, but significant, driving force that sustains the crystal's contraction;
    \item above 70 K: the remaining higher-frequency modes are characterized by predominantly positive Grüneisen parameters. Above roughly 50 K, their heat capacity begins to rise sharply. By approximately 70 K, the massive positive contribution from these densely populated higher-energy branches fully compensates, and then overtakes, the negative contributions from the TA and TO modes.
\end{itemize}

This mode-resolved quantitative breakdown confirms that the cryogenic NTE in ice I$_c$ is a sequential consequence of the early thermal excitation of negative-$\gamma$ TA and TO modes.

\section*{References} 

\bibliography{pnas-sample}